\documentclass[12pt]{article}

\usepackage{sbc-template}
\usepackage{graphicx,url}
\usepackage[utf8]{inputenc}
\usepackage{listings} 
\usepackage{fancybox}
\usepackage{xcolor}
\usepackage{color}
\usepackage{amsmath}
\usepackage{tcolorbox}
\usepackage{booktabs}
\usepackage{soul}
\usepackage{pifont}

\title{Refactoring Assertion Roulette and Duplicate Assert\\ test smells: a controlled experiment}

\author{Railana Santana\inst{1}, Luana Martins\inst{1}, Tássio Virgínio\inst{2},  \\ Larissa Soares\inst{1,3}, Heitor Costa\inst{4}, Ivan Machado\inst{1}}

\date{\today}

\address{
    Federal University of Bahia (UFBA) -- Salvador, BA -- Brazil
    \vspace{-10pt}
    \nextinstitute
    Federal Institute of Tocantins (IFTO) -- Paraíso do Tocantins, TO -- Brazil
    \vspace{-10pt}
    \nextinstitute
    State University of Feira de Santana (UEFS) -- Feira de Santana, BA -- Brazil
    \vspace{-10pt}
    \nextinstitute
    Federal University of Lavras (UFLA) -- Lavras, MG -- Brazil
    \email{railana.santana@ufba.br, martins.luana@ufba.br,}
    \vspace{-10pt}
    \email{lrsoares@uefs.br, tassio.virginio@ifto.edu.br,}
    \vspace{-10pt}
    \email{ heitor@ufla.br, ivan.machado@ufba.br}
}

\usepackage{listings}
\usepackage{fancybox}
\usepackage{tikz}
\usetikzlibrary{tikzmark}

\definecolor{verde}{rgb}{0.25,0.5,0.35}
\definecolor{jpurple}{rgb}{0.5,0,0.35}
\definecolor{vermelho2}{rgb}{13.0, 8.0, 14.0}
\definecolor{amarelo}{rgb}{1.0, 0.98, 0.8}

\newcommand{\hll}[1]{{\sethlcolor{amarelo}\hl{#1}}}
\newcommand{\correcoes}[1]{{\color{black}#1}}

\newcommand{\citacao}[1]{{\color{black} #1}}
\begin{document} 
\maketitle

\def \questionOne {How does \texttt{RAIDE} facilitate the AR and DA test smells detection compared to \texttt{tsDetect}? }
 
\def \questionTwo {How does \texttt{RAIDE} facilitate the smelly test code refactoring with the AR and DA test smells compared to manual refactoring?}

\begin{abstract}
Test smells can reduce the developers' ability to interact with the test code. Refactoring test code offers a safe strategy to handle test smells. However, the manual refactoring activity is not a trivial process, and it is often tedious and error-prone. This study aims to evaluate \texttt{RAIDE}, a tool for automatic identification and refactoring of test smells. We present an empirical assessment of \texttt{RAIDE}, in which we analyzed its capability at refactoring Assertion Roulette and Duplicate Assert test smells and compared the results against both manual refactoring and a state-of-the-art approach. The results show that \texttt{RAIDE} provides a faster and more intuitive approach for handling test smells than using an automated tool for smells detection combined with manual refactoring.
\end{abstract}

\section{Introduction}\label{section:introduction}

Writing automated tests requires more than just understanding the business rules implemented in the source code, as the test engineer should be skilled enough to build well-structured test cases. In addition, the complexity of the system under test aligned with the lack of knowledge and experience may lead test engineers to use bad practices to either design or implement the test code \cite{garousi2018}.

Bad design practices in the test code are commonly referred to as test smells \cite{van2001refactoring}. Test smells have recently gained importance given their effects on the performance of software testing activities, especially from a maintenance perspective \cite{Virginio2020, mapping_testsmells}. For instance, \textit{Empty Test} is a test smell that occurs when a test method does not contain any executable instructions. Since the method does not have a body, the test always passes. When developers introduce behavior-breaking changes, an empty test does not notify alternated outcomes \cite{van2001refactoring, peruma2019distribution}. 

While good testing practices and guidelines can prevent test smells, test engineers do not always follow them. In particular, whether a software project already comprises a large set of tests, it may not be cost-effective to create novel tests from scratch. An alternative is to employ test-specific refactoring strategies to improve the test code quality without changing its behavior \cite{van2001refactoring}. Due to the lack of test suites aimed to test themselves, there is a need for automated tools to refactor the test code and keep its behavior.

The literature has introduced a small set of automated tools to refactor test code \cite{mapping_testsmells}. For example, \texttt{RTj} \cite{Martinez2020} is a command-line tool that supports detecting and refactoring Rotten Green Tests, i.e., a test that passes during execution but has assertions rarely executed. \texttt{DARTS} \cite{Lambiase2020} is an IntelliJ plugin that supports detecting and refactoring three test smells (General Fixture, Eager Test, and Lack of Cohesion of Test Methods). These tools address a small number of test smells when using the JUnit framework to develop test cases. \citacao{\cite{mapping_testsmells}} point out that the existing tools do not provide details concerning the accuracy of their refactoring capabilities or usability. To expand the set of refactoring strategies for test smells, we previously presented a systematic approach to detect and refactor two test smells: \textit{Assertion Roulette} (AR) and \textit{Duplicate Assert} (DA). In addition, we implemented \texttt{RAIDE}, automated tool support for test smell refactoring \cite{santana2020raide}. \correcoes{\texttt{RAIDE} is an open-source tool with a user-friendly interface that can detect and refactor test smells with just a few clicks.} 

In this paper, we present an empirical study to evaluate \texttt{RAIDE}. We aimed to answer the Research Question: \textbf{\textit{How does \texttt{RAIDE} support users to detect test smells and refactor the test code?}} As the related tools do not support refactoring strategies for the AR and DA test smells, we compared \texttt{RAIDE} with manual refactoring. We asked twenty test engineers to refactor test code from two projects. While using \texttt{RAIDE}, the test engineers had access to an interface integrated into the Eclipse IDE. They could detect and refactor the test smells. Otherwise, for the manual refactoring, the test engineers used the \texttt{tsDetect} to detect the test smells. This is a state-of-the-art test smell detection tool \cite{peruma2019distribution}, Given that \texttt{tsDetect} does not allow automated refactoring, the participants had to use their strategies to refactor the test smells. 

In summary, we contributed with a controlled experiment to compare \texttt{RAIDE} with one state-of-the-art approach and discuss the usefulness of an IDE-integrated tool that automatically detects and refactors test smells. Controlled experiment findings can support the community of researchers and developers in building and maintaining intuitive tools to detect and refactor test code automatically. Besides, we collected the participants' perceptions of the tools, e.g., the \texttt{RAIDE} and \texttt{tsDetect} limitations, a feedback key to the continuity and evolution of these tools.

The remainder of this paper is structured as follows. Section \ref{section:test_smells} introduces the concept of test smells and the \texttt{RAIDE} and \texttt{tsDetect} tools. Section \ref{section:controlled_experiment} details the experiment design and the main results. Section \ref{section:results} discusses the evaluation results. Section \ref{section:treats_to_validaty} presents the threats to the validity. Section \ref{section:related_work} presents related work. Section \ref{section:conclusion} concludes the paper.
\section{Background}
\label{section:test_smells}

Test smells result from bad design choices implemented in the test code \cite{greiler2013strategies}. Smells in test code can affect its quality, mainly understandability and maintainability. Consequently, test smells can reduce the effectiveness of test cases to detect faults and the developers' ability to interact with the test code \cite{garoussi2015}. 

Although there are several test smells, some of them are more prevalent. \citacao{\cite{palomba2016diffusion}} empirically evaluated the diffusion of test smells in automatically generated JUnit test classes in 110 open-source software projects. The results showed that 83\% of those classes are affected by at least one test smell. The most frequent test smells were the \textit{AR} (54\%) and \textit{Test Code Duplication} (33\%). In our study, we considered \textit{DA}, once it is a representative type of \textit{Test Code Duplication} test smell. 

In addition, \citacao{ \cite{peruma2018smell}} conducted a large-scale empirical study on the test smells occurrence, distribution, and impact in the maintenance of open-source Android applications. They also observed that the \textit{AR} test smell occurred in more than 50\% of the test classes. In a multivocal literature review, \citacao{ \cite{garousi2018}} reported the most extensive catalog of test smells and a summary of guidelines, techniques, and tools to handle test smells. The authors pointed out that test smells related to code duplication and code complexity (test redundancy and long test, respectively) have been the most discussed ones in the literature.

The recurring number of studies reporting on the occurrence of the \textit{AR} test smell \cite{peruma2018smell, palomba2016diffusion} and the repercussion on test smells related to code duplication \cite{garousi2018} led us to investigate those two test smells and then propose \texttt{RAIDE}. We next introduce these test smells.

\vspace{-0.2cm}
\subsection{Assertion Roulette (AR)}

\correcoes{In JUnit, \textit{assertions} have an optional first argument of the String type to explain what each \textit{assertion} is testing.} AR occurs when a test method has several undocumented \textit{assertions}, making understanding difficult during maintenance and challenging to detect the \textit{assertion} if the method fails. Listing \ref{lst:ExemploAR} presents a code excerpt with multiple \textit{assertions} without a return message (lines 93 to 95). The example presents a method from the \texttt{TestAbstractPartial.java}\footnote{\textcolor{black}{Available at} \url{https://bit.ly/35Q56KV}} test class of the \textbf{Joda-Time} project.

\begin{lstlisting} [float=ht, caption = {Test code with the Assertion Roulette test smell}, label={lst:ExemploAR}, firstnumber = 90]
public void testGetValues() throws Throwable {
    MockPartial mock = new MockPartial();
    int[] vals = mock.getValues();
    assertEquals(2, vals.length);
    assertEquals(1970, vals[0]);
    assertEquals(1, vals[1]);
}
\end{lstlisting}

\noindent
\textbf{Possible Effect:} Multiple \textit{assertion} statements in a test method without a descriptive message can affect test readability, comprehensibility, and maintainability. Multiple \textit{assertions} make it difficult to detect which assertion gave an error in a test failure.

\noindent
\textbf{Detection:} To check if a test method has \textit{assertions} without explanation/message (parameter in the \textit{assertion} method).

\noindent
\textbf{Refactoring:} To include \textit{assertion} explanations in each \textit{assertion}. Listing \ref{lst:ExemploARRefactored} shows the Listing \ref{lst:ExemploAR} code refactored with the appropriate explanations for each assert (text highlighted in yellow).

\begin{lstlisting} [float=ht, caption = {Test code after refactoring the Assertion Roulette test smell}, label={lst:ExemploARRefactored}, firstnumber = 90]
public void testGetValues() throws Throwable {
    MockPartial mock = new MockPartial();
    int[] vals = mock.getValues();
    assertEquals((*@\textcolor{blue}{\hll{"Vals size 2"}}@*), 2, vals.length);
    assertEquals((*@\textcolor{blue}{\hll{"Year Equal 1970"}}@*), 1970, vals[0]);
    assertEquals((*@\textcolor{blue}{\hll{"Month 1"}}@*), 1, vals[1]); 
}
\end{lstlisting}

\vspace{-0.2cm}    
\subsection{Duplicate Assert (DA)}

DA occurs when a test method tests the same condition multiple times in the same test method \cite{peruma2019distribution}. Listing \ref{lst:ExampleDA} shows a code excerpt with two \textit{assertions} with the same parameters (lines 361 and 363). The example presents one method from the \texttt{TestPeriodFormatterBuilder.java}\footnote{\textcolor{black}{Available at} \url{https://bit.ly/3oriGL1}} test class of the \textbf{Joda-Time} project.

\begin{scriptsize}
\begin{lstlisting} [float=ht, caption = {Test code with the Duplicate Assert test smell}, label={lst:ExampleDA}, firstnumber=356]
public void testPluralAffixParseOrder() {
    PeriodFormatter f = builder.appendDays()
        .appendSuffix("day", "days").toFormatter();
    String twoDays = Period.days(2).toString(f);
    Period period = f.parsePeriod(twoDays);
    assertEquals(Period.days(2), period);
    period = f.parsePeriod(twoDays.toUpperCase(Locale.ENGLISH));
    assertEquals(Period.days(2), period);
}
\end{lstlisting}
\end{scriptsize}

\noindent \textbf{Possible Effect:} That test smell hinders test readability and maintenance, as there are repeated assertions (with the same parameters) without explaining the purpose/objective of the test method. In general, DA creates a scenario that violates the responsibility of each method to fulfill a single objective.

\noindent
\textbf{Detection:} To check if the test method contains two or more \textit{assertion} statements with the same parameters.\\
\textbf{Refactoring:} To create one test method for testing the same condition with different values. Listing \ref{lst:ExampleDARefactored} shows a code excerpt from Listing \ref{lst:ExampleDA} refactored (text highlighted in yellow), which extracts the duplication for a new method (\texttt{testPluralAffixParseOrderExtracted}).

\begin{scriptsize}
\begin{lstlisting} [float=ht, caption = {Test code after refactoring the Duplicate Assert test smell}, label={lst:ExampleDARefactored}, firstnumber=356]
public void testPluralAffixParseOrder() {
    PeriodFormatter f = builder.appendDays().
        appendSuffix("day", "days").toFormatter();
    String twoDays = Period.days(2).toString(f);
    Period period = f.parsePeriod(twoDays);
    assertEquals(Period.days(2), period);
}

(*@\textcolor{verde}{\hll{/*  Extracted Method  */}}@*)
(*@\textcolor{jpurple}{\textbf{\hll{public void}}}@*)(*@\hll{ testPluralAffixParseOrderExtracted() \{ }@*)
    (*@\hll{PeriodFormatter f = builder.appendDays(). }@*)
         (*@\hll{appendSuffix(}@*)(*@\textcolor{blue}{\hll{"day"}}@*)(*@\hll{,}@*)(*@\textcolor{blue}{\hll{"days"}}@*)(*@\hll{).toFormatter(); }@*)
    (*@\hll{String twoDays = Period.days(2).toString(f); }@*)
    (*@\hll{Period period = f.parsePeriod(twoDays.toUpperCase(Locale.ENGLISH));}@*)
    (*@\hll{assertEquals(Period.days(2), period); }@*)
 (*@\hll{\} }@*)
\end{lstlisting}
\end{scriptsize}
\vspace{-0.2cm}
\subsection{\texttt{tsDetect}}

\begin{figure}[htb]
     \centering
     \includegraphics[width=0.65\columnwidth]{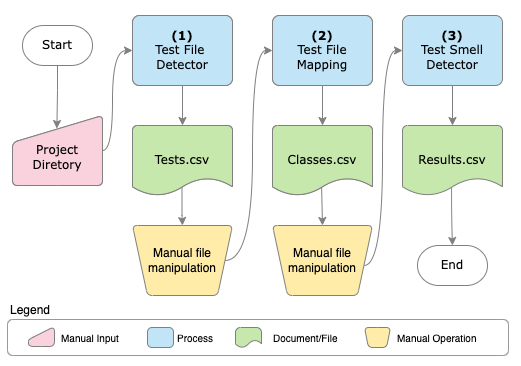}
     \caption{Process for running the \texttt{tsDetect} tool}
     \label{img:processo_tsd}
     \vspace{-2pt}
\end{figure}

\texttt{tsDetect} has been reported in a recently published literature review as a comprehensive tool for detecting test smells in Java projects \cite{mapping_testsmells}. The tool covers 19 test smells with a precision score ranging from 85\% to 100\% and a recall score from 90\% to 100\% in open-source Android apps \cite{peruma2019distribution}. Given those precision and recall scores, researchers built \texttt{tsDetect}-based tools \cite{jnose,kim2021secret}.
\texttt{tsDetect} tool indicates the existence of test smells in a test class based on a three-step detection process (Figure \ref{img:processo_tsd}): 1) \textit{Test File Detector} - reads the project test files; 2) \textit{Test File Mapping} -  links the test files to the production files under test; and 3) \textit{ Test Smell Detector} - detects smells in test code.

\vspace{-2pt}

In step (1), \textit{Test File Detector} generates the \texttt{Tests.csv} file, which contains the path of the test classes from one software project. That file is input to \textit{Test File Mapping}. Next, step (2) establishes the relationship between the test and production classes. It creates the \texttt{Classes.csv} file, which contains the project name, the path of each test class, and the production classes. That file is input to \textit{Test Smell Detector}. Step (3) is responsible for analyzing test smells for the project based on the \texttt{Classes.csv} file. The output is the \texttt{Results.csv} file, which indicates the presence (true) or absence (false) of test smells in the test classes. 

\vspace{-0.2cm}
\subsection{\texttt{RAIDE}}
\label{section:raide_tool}

\texttt{RAIDE} is an AST (Abstract Syntax Tree)-based tool developed as an Eclipse open-source plugin to detect and refactor test smells \cite{santana2020raide}. We reused rule-based components from \texttt{tsDetect} and performed improvements to detect test smells. Whereas \texttt{tsDetect} works as a command-line tool that indicates the presence of test smells, \texttt{RAIDE} has a user-friendly Graphical User Interface (GUI), which identifies and indicates the exact location (code lines) of the AR and DA test smells. Besides, it includes one feature for automated refactoring of those test smells.

For the test code refactoring activity to succeed, the detection must be precise and explicit, pointing to the source code line where the test smell is located. However, not all tools report the exact location of test smells. For instance, the \texttt{tsDetect} only informs whether a class is affected by a test smell. Thus, the users need to analyze the entire test class to identify the test smells and refactor them. Conversely, \texttt{RAIDE} exhibits the  test smells exact location for users and provides a user-friendly GUI.

\texttt{RAIDE} uses graphical components from \texttt{JDeodorant}\footnote{\textcolor{black}{Available at} \url{https://github.com/tsantalis/JDeodorant}}, an Eclipse plugin, to detect and refactor code smells in java code and reuses the components \texttt{tsDetect} tool: i) AST of the project, responsible for detecting test classes and code structure; and ii) AR test smell detection rules. In addition, we improved the detection rules to meet the scenario with an empty string (``'') or space string (`` '') in the explanation parameter and inform the line affected by the test smell. The way \texttt{tsDetect} implements the DA test smell detection does not allow code reuse for accurately detecting each test smell in \texttt{RAIDE}. Therefore, we built modules from scratch in \texttt{RAIDE} to detect and refactor the DA test smell and to refactor the AR test smell. Some limitations of \texttt{RAIDE} include: i) the implementation of detection rules is based on JUnit 4, other JUnit versions may require adaptations in such rules; and ii) the tool's execution detects the \textit{Assertion Roulette} or \textit{Duplicate Assert}, not both at the same time.

\begin{figure}[htb]
     \centering
     \includegraphics[width=0.65\columnwidth]{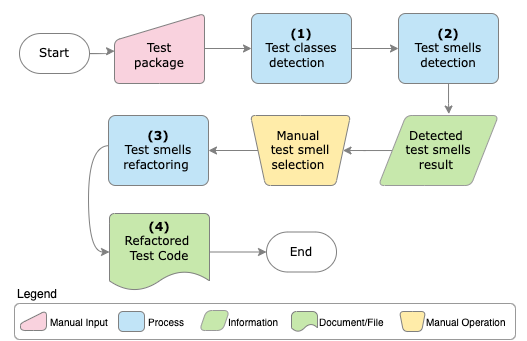}
     \caption{Process for running the \texttt{RAIDE} tool}
     \label{img:processo_raide}
     \vspace{-2pt}
\end{figure}

Figure \ref{img:processo_raide} shows the \texttt{RAIDE} tool process to identify and refactor test smells. The user should provide as input: the \textbf{test package} of the project under analysis, and the test smell the plugin should detect and refactor. In step (1), the \textbf{test classes detection} identifies all JUnit classes in the \textbf{test package}. In step (2), the \textbf{test smells detection} detects a specific type of test smell and presents the \textbf{test smell detection results} in an Eclipse view. In step (3), \textbf{manual test smells selection} requires the user intervention to select which test smell instances(s) he would like to refactor. Then, in step (4), \texttt{Test smells refactoring} shows the user how to refactor the code, and the user can take the decision of accepting the \textbf{refactored test code}.

\vspace{-2pt}

\begin{figure}[htb]
     \centering
     \includegraphics[width=0.8\columnwidth]{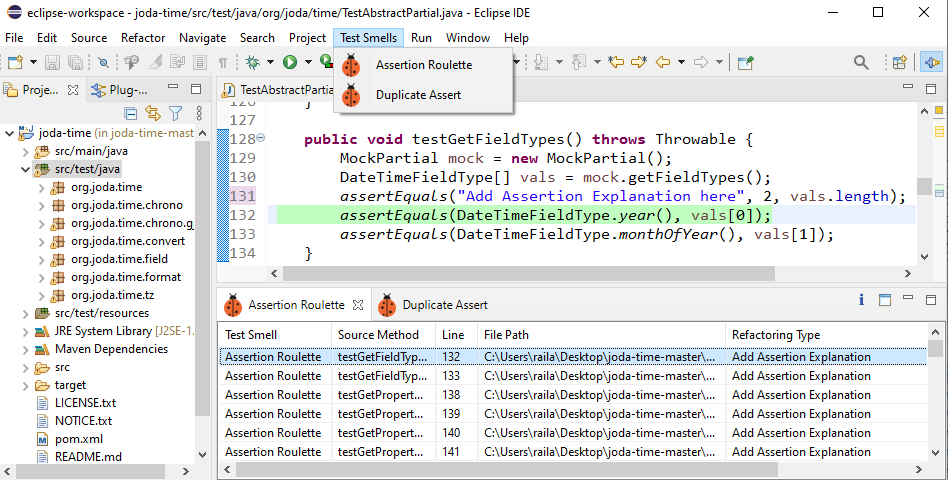}
     \caption{Screenshot of the \texttt{RAIDE} plugin under execution}
     \label{img:raide}
     \vspace{-2pt}
\end{figure}

Figure \ref{img:raide} shows a screenshot of the Eclipse IDE with the \texttt{RAIDE} plugin running on the Joda-Time\footnote{Available at \url{https://github.com/JodaOrg/joda-time}} project. \texttt{RAIDE} refactored line 131 of method \texttt{testGetFieldTypes()} after detecting the AR test smell. \texttt{RAIDE} included the explain parameter ``Add Assertion Explanation here'' to correct that test smell. The user must replace the default string with an explanatory message about the assertion to remove the AR test smell from the code. In Figure \ref{img:raide}, it is also possible to see that \texttt{RAIDE} also detected the AR test smell on line 132. After double-clicking on the detected test smell, the tool redirects the user to the highlighted line.

\vspace{-2pt}

\vspace{-4pt}
\section{Empirical Assessment}
\label{section:controlled_experiment}

In this study, we carried out a controlled experiment to evaluate how the automated process proposed by \texttt{RAIDE} assists the test engineers to (i) detect test smells and (ii) refactor test code. To answer the main research question (RQ), we split it into two sub-RQs: \textbf{RQ\textsubscript{1})} \questionOne \textbf{RQ\textsubscript{2})} \questionTwo \ For RQ\textsubscript{1}, we compared \texttt{RAIDE} and \texttt{tsDetect} concerning how both detect test smells and show the results. For RQ\textsubscript{2}, we compared refactorings with \texttt{RAIDE} to manual refactorings. Thus, we measured the participants' time taken to identify the test smells in the test code with the support of \texttt{RAIDE} and \texttt{tsDetect}. After, we measured the refactorings time performed manually and with \texttt{RAIDE}.

\begin{figure}[t]
    \centering
    \includegraphics[width=0.8\linewidth]{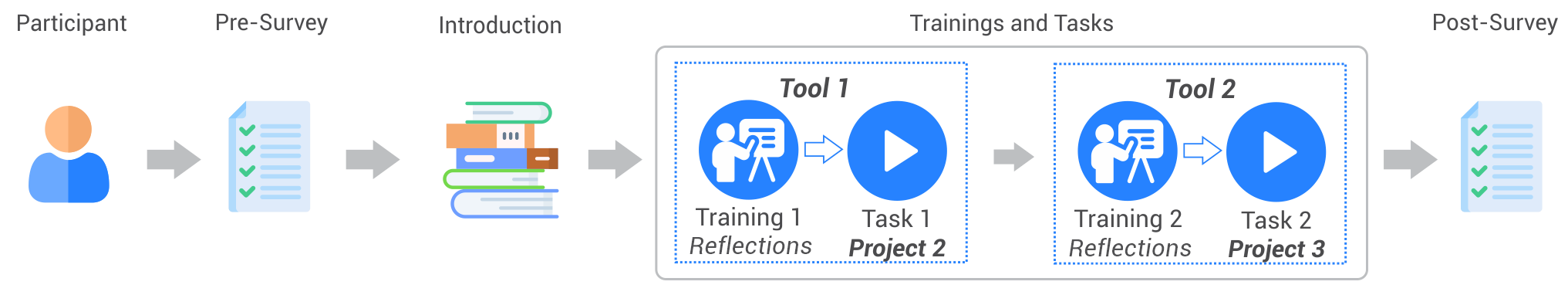}
    \caption{Experiment Flow}
    \label{fig:design}
    \vspace{-2pt}
\end{figure}

\textbf{Experiment Overview.} Before joining the experiment session, the participants filled out an online form with questions regarding their experience in software programming and testing, Java language, JUnit framework, Eclipse, and test smells. Next, we presented the concepts about test smells (AR and DA, particularly) and detection and refactoring processes. We also informed the objective of the experiment. Figure \ref{fig:design} shows an overview of the experiment steps. The participants performed two tasks, one for \texttt{RAIDE} and the other for \texttt{tsDetect}. Each task encompassed the analysis of a different software project. Before running the experiment, the participants received training on using and executing the tool shortly after. We used the same project in the training sessions of both tools (\textit{Project 1} - Figure \ref{fig:design}). After training, the participants performed the actual experiment tasks, as follows: (i) they used \texttt{RAIDE} and \texttt{tsDetect} to detect two test smells (AR and DA) and informed the location (code lines) of the test smells, and (ii) they refactored the code to remove the two test smells. Tasks 1 and 2 are similar, but the participants used different tools (either \textit{Tool 1} or \textit{Tool 2}) and projects (either \textit{Project 2} or \textit{Project 3}) (Figure \ref{fig:design}). In the end, the participants answered an online post-survey\footnote{\textcolor{black}{Available at} \url{https://zenodo.org/record/5978022#.Yf55hOrMKUk}}, comprising questions about the experiment execution and their perception of each tool.

\vspace{-2pt}

\textbf{Pilot Study.} We performed a pilot study with five participants, two graduate students, two practitioners, and another acting in both roles. We found that they took 47 seconds, on average, using \texttt{RAIDE} and 15 minutes, on average, using \texttt{tsDetect} to detect the test smells. The pilot study was critical in reviewing the concepts and standardizing the training, especially the commands needed to use \texttt{tsDetect}. It also was helpful to assess whether the participants could understand the tasks and tools. Data gathered in this pilot study was not considered in the final analysis.

\textbf{Participants.} We recruited twenty participants, where ten participants (50\%) were from the academy, five participants (25\%) were from industry, and five participants (25\%) acted in both roles. All participants held at least a B.Sc. degree, seven participants (35\%) were M.Sc. students, and seven participants (35\%) were Ph.D. students. They come from sixteen different Brazilian institutions, eight different universities (U1 to U8), and eight different software development companies (C1 to C8). Some of them also belong to more than one institution (P1, P7, P12, P15, P17, P18, and P20). In addition, they have different roles: Agile Coach (AC), Developer (D), Database Admin (DA), Lecturer (L), Researcher (R), and Requirements Analyst (RA).
Most participants (80\%) had already heard about test smells. Six participants (30\%) had specific knowledge about test smell in either industry or academia. Table \ref{tab:participantes_experiencia} shows detailed information. Although the participant's profile was collected, we did not investigate the relationship between their experience and performance in carrying out the tasks.

\begin{table}[th]
\scriptsize
\def \arraystretch{1.1}
\setlength{\tabcolsep}{0.05cm}
\centering
\caption{Participants' profile and experience (in years)}
\begin{tabular}{ccccccccccc}
 \toprule
\textbf{ID} & \textbf{Group} & \textbf{Education} & \textbf{Profile} &\textbf{Institution} & \textbf{Programming} & \textbf{JAVA} &\textbf{Eclipse} & \textbf{Testing} & \textbf{JUnit} & \textbf{Test Smells}\\ 
 \midrule
P01	&	1	&	Ph.D. student	&	R/L	&	U1/U2	&	10+	&	10+	&	10+	&	10+	&	1$\vdash$ 5 & \ding{51}\\
P02	&	2	&	M.Sc. student	&	R	&	U1	&	10+	&	0-1	&	0-1	&	5$\vdash$ 10	&	0-1 & \ding{85}\\
P03	&	3	&	Ph.D. student	&	R	&	U1	&	0-1	&	0	&	1$\vdash$5	&	1 $\vdash$5	&	1$\vdash5$ & \ding{51}\\
P04	&	4	&	B.Sc.	&	RA	&	C1	&	1$\vdash$5	&	1$\vdash$5	&	1$\vdash$5	&	1$\vdash$5	&	0-1 & \ding{55}\\
P05	&	1	&	B.Sc.	&	D	&	C2	&	5$\vdash$10	&	1$\vdash$5	&	0-1	&	0-1	&	0-1 & \ding{51}\\
P06	&	2	&	M.Sc. student	&	R/L	&	U1	&	5$\vdash$10	&	1$\vdash$5	&	1$\vdash$5	&	1$\vdash$5	&	1$\vdash$5 & \ding{51}\\
P07	&	3	&	Ph.D. student	&	R/L	&	U1/U3	&	1$\vdash$5	&	1$\vdash$5	&	1$\vdash$5	&	5$\vdash$10	&	0-1 & \ding{72}\\
P08	&	4	&	B.Sc.	&	L	&	U4	&	5$\vdash$10	&	1$\vdash$5	&	1$\vdash$5	&	0-1	&	0 & \ding{55}\\
P09	&	1	&	B.Sc.	&	D	&	C1	&	1$\vdash$5	&	1$\vdash$5	&	0	&	0-1	&	0-1 & \ding{51}\\
P10	&	2	&	B.Sc.	&	D	&	C3	&	1$\vdash$5	&	0-1	&	0-1	&	0-1	&	0 & \ding{55}\\
P11	&	3	&	M.Sc. student	&	R/DA	&	U5	&	1$\vdash$5	&	1$\vdash$5	&	1$\vdash$5	&	0	&	0 & \ding{85}\\
P12	&	4	&	Ph.D. student	&	R/L	&	U1/U6	&	5$\vdash$10	&	1$\vdash$5	&	1$\vdash$5	&	0-1	&	0-1 & \ding{72}\\
P13	&	1	&	Ph.D. student	&	R	&	U7	&	10+	&	1$\vdash$5	&	1$\vdash$5	&	0	&	0 & \ding{72}\\
P14	&	2	&	Ph.D. student	&	R	&	U1	&	1$\vdash$5	&	1$\vdash$5	&	0-1	&	0-1	&	0 & \ding{73}\\
P15	&	3	&	M.Sc. student	&	R/D	&	U1/C4	&	1$\vdash$5	&	1$\vdash$5	&	0-1	&	0	&	0 & \ding{51}\\
P16	&	4	&	B.Sc.	&	D	&	C5	&	5$\vdash$10	&	1$\vdash$5	&	1$\vdash$5	&	1$\vdash$5	&	1$\vdash$5  & \ding{51}\\
P17	&	1	&	M.Sc. student	&	R/AC	&	U5/C6	&	5$\vdash$10	&	5$\vdash$10	&	5$\vdash$10	&	0-1	&	0-1 & \ding{51}\\
P18	&	2	&	M.Sc. student	&	R/D	&	U1/C7	&	5$\vdash$10	&	1$\vdash$5	&	1$\vdash$5	&	0	&	0 & \ding{72}\\
P19	&	3	&	Ph.D. student	&	R	&	U8	&	10+	&	10+	&	10+	&	10+	&	10+ & \ding{51}\\
P20	&	4	&	M.Sc. student	&	R/D	&	U1/C8	&	1$\vdash$5	&	1$\vdash$5	&	1$\vdash$5	&	0	&	0 & \ding{55}\\
 \bottomrule

\multicolumn{11}{c}{Labels: (\ding{55}) Never heard about them; (\ding{51}) Already heard anything about them, but had never worked with them;} \\
\multicolumn{11}{c}{(\ding{73}) Knew a little bit about them; (\ding{72}) Heard about test smells and had already worked with them;} \\
\multicolumn{11}{c}{and (\ding{85}) Researcher investigating the topic of test smells.} 
\end{tabular}
\label{tab:participantes_experiencia}
\end{table}

\textbf{Experiment Material and Tasks.} We selected three open-source projects for this experiment: \textit{Reflections} project for training and \textit{Joda-Time} and \textit{Commons-collections} projects for executing tasks 1 and 2 (\textit{Project 2} and \textit{Project 3} - Figure \ref{fig:design}, but not necessarily in that order). We chose the projects considering the limitations of \texttt{RAIDE} and \texttt{tsDetect}, which support Maven projects and tests with JUnit (version 4). Due to the size of the projects, we decided to consider only two test classes for the experiment. Therefore, the \textit{Reflections}, \textit{Joda-Time}, and \textit{Commons-collections} projects would have a similar complexity level, number of methods, and number of test smells. Regarding the test smells in the projects, one method has the AR test smell (4 \textit{assertions} without explanation), and another one has two pairs of the DA test smell (4 \textit{assertions} in the same method).

\textbf{Design and Procedure.} In our experiment, we used a crossover design \cite{vegas2015crossover} to avoid the learning effect, as the participants performed two tasks in a row. The projects and the tools are the independent variables, and time is the dependent variable of the experiment. During the experiment with each participant, we captured the audio and the computer screen to count later the time spent by them to detect and refactor the test smells in each task of the experiment. When a participant inaccurately identified or refactored the test smell, the researcher reported that the task had not been completed yet, and the time continued counting until the task be completed correctly.

\textbf{Data Analysis.} We performed the Shapiro-Wilk test, with a significance level of 5\%, to verify the data distribution for the data analysis. As a result, the data distribution is not normal. Then, we selected the Mann-Whitney paired test \cite{mann1947test}, with a significance level of 5\%, to answer RQ\textsubscript{1} and RQ\textsubscript{2} (sub-RQs). We defined the null hypothesis  to investigate the RQ\textsubscript{1}: \textit{The detection time of AR and DA test smells with \texttt{RAIDE} is similar to detection with \texttt{tsDetect}.} Regarding the RQ\textsubscript{2}, we defined the null hypothesis: \textit{The refactoring time of AR and DA test smells with \texttt{RAIDE}  is similar to manual refactoring.} 
\section{Results and Discussion}\label{section:results}

This section presents and discusses the results gathered from the empirical assessment.

\subsection{Detection of Test Smells (RQ1)}\label{sec:RQ1_answer}

We collected and analyzed the time spent by the participants to complete the task of detecting and locating the AR and DA test smells with \texttt{RAIDE} and \texttt{tsDetect} to answer RQ\textsubscript{1}. The participants completed that task in 63.95 seconds (s) on average with a standard deviation (\textit{sd}) of 28.12s using \texttt{RAIDE} and 679.20s on average with \textit{sd} = 248.71s using \texttt{tsDetect}. Thus, those values suggest that detecting and locating the AR and DA test smells with \texttt{tsDetect} is slower than with \texttt{RAIDE}.

The difference between \texttt{RAIDE} and \texttt{tsDetect} is the first one to be an IDE-integrated tool, in which participants select the test smell they want to analyze and report the lines with the smells highlighted by the tool. In \texttt{tsDetect}, the participants need to run three different tools, open a \texttt{.csv} file to check which classes have test smells, and manually inspect the source code. Therefore, the participants spend most of the time using the \texttt{Test File Detector} and \texttt{Test File Mapping} tools and performing adjustments in the \texttt{.csv} files. Although the tools are similar in execution time, they present a wide difference in their efficiency regarding how fast they place the user in front of the test smells.

\begin{figure}[th]
    \centering
    \includegraphics[width=0.9\linewidth]{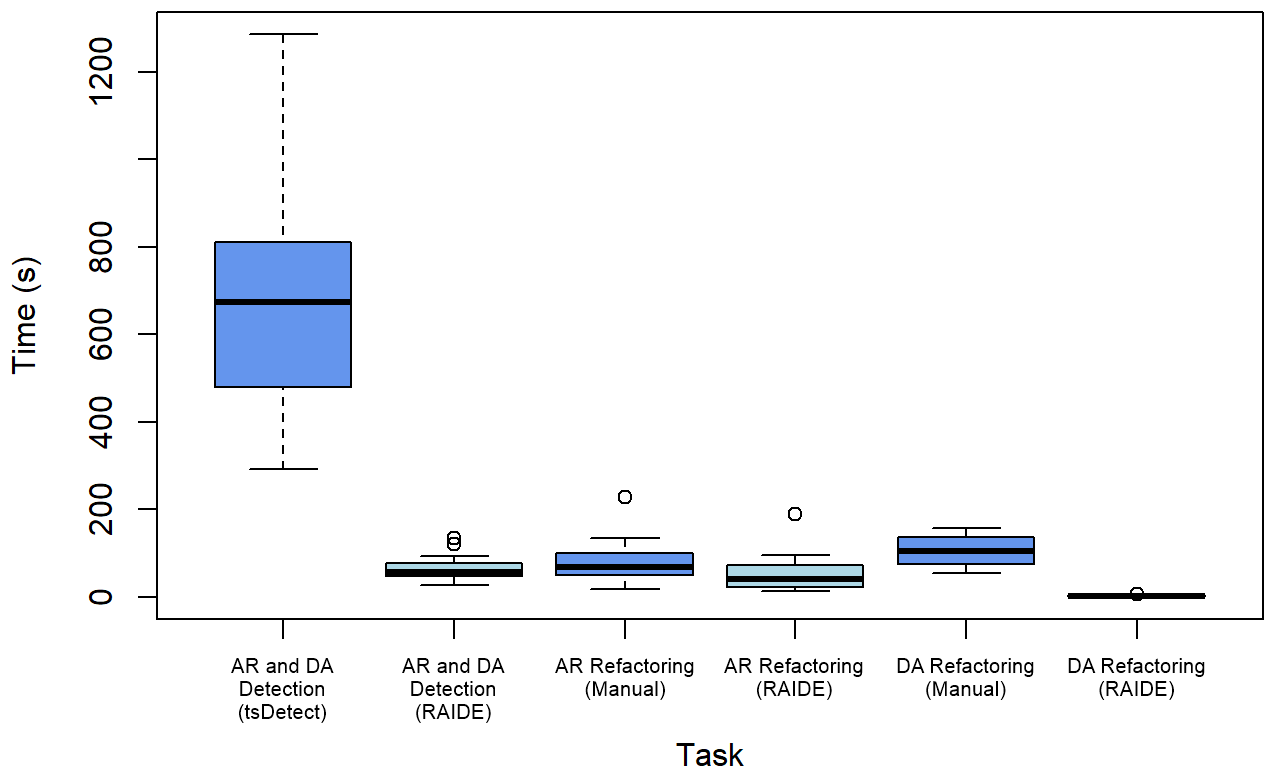}
    \caption{Comparison of the test smell detection and refactoring times between \texttt{RAIDE} and \texttt{tsDetect}}
    \label{fig:box_comparison}
    \vspace{-2pt}
\end{figure}

We also analyzed whether there is a statistically significant difference in identifying the AR and DA test smells. Since our data did not have a normal distribution (\textit{p-value} = 7.748e-05), we performed the non-parametric Mann-Whitney test. The test indicated a significant difference (\textit{p-value} = 9.537e-07) between the average time spent using \texttt{tsDetect} and \texttt{RAIDE}. Thus, we refuted the null hypothesis (\textit{H\textsubscript{0}1}) and answered RQ\textsubscript{1}. We also observed that data resulting from the time measures with \texttt{tsDetect} is more dispersed than the time using \texttt{RAIDE}. Figure \ref{fig:box_comparison} shows boxplots on a logarithmic scale comparing data gathered from \texttt{RAIDE} and \texttt{tsDetect}. There is no overlap between the boxes of \texttt{RAIDE} and \texttt{tsDetect}, which means a difference between them. In addition, the longest reported time for detecting the AR and AD test smells with \texttt{RAIDE} was much shorter than the shortest time to detect them with \texttt{tsDetect}. The participants using \texttt{RAIDE} achieved similar results in terms of efficiency. However, we can not say the same when they used \texttt{tsDetect}, which indicates that \texttt{RAIDE} standardizes how participants detect the AR and AD test smells, regardless of their experience.

\vspace{-2pt}

Our findings show that all participants encountered the AR and DA test smells faster with \texttt{RAIDE} than using \texttt{tsDetect}. The mean difference between the time of each task was more than 615s, considering all the steps needed to run \texttt{tsDetect}. Moreover, the participants pointed out many advantages that \texttt{RAIDE} would have over \texttt{tsDetect} related to identifying the AR and DA test smells. For example, according to P02, \texttt{RAIDE} stood out because \textit{``The learning process is easier and faster, it has fewer steps to take. It integrates into a development tool with a graphical interface, facilitating its use. It also identifies the lines that each test smell occurs and suggests refactorings.''}

\begin{tcolorbox}
\textbf{Summary:} The AR and DA test smells could be identified ten times faster with \texttt{RAIDE} than with the state-of-the-art tool. \texttt{RAIDE} offers a better user experience, e.g., it enables detecting test smells with just one click and highlights their location in the source code. 
\end{tcolorbox}

\subsection{Test Code Refactoring (RQ2)}

We collected and analyzed the time spent by the participants to refactor test code with the AR and DA test smells using \texttt{RAIDE} and \texttt{tsDetect}.

\noindent
\textbf{AR Refactoring.} The participants could refactor the test code with \texttt{RAIDE} on an average of 49.9s (\textit{sd} = 41.35s), while the manual refactoring took 78.30s (\textit{sd} = 46.39s). As the refactoring of \texttt{RAIDE} is by line, we need to count the time spent by the user to select each test smell of the method correctly and refactor them individually. For the statistical analysis, we performed the Shapiro-Wilk normality test. We found that the refactoring time of the AR test smell does not have a normal distribution (\textit{p-value} = 6.607e-05). Therefore, we also used the Mann-Whitney test. Although the refactoring with \texttt{RAIDE} has shown similar dispersion to manual refactoring, the results indicated a significant difference between automated refactoring and manual refactoring (\textit{p-value} = 0.022). It leads us to understand that \texttt{RAIDE} has a shorter detection time, which refutes the null hypothesis (\textit{H\textsubscript{0}2}). Figure \ref{fig:box_comparison} shows boxplots with such comparisons, the median line of the box of \texttt{RAIDE} is outside the box of manual refactoring, which confirms a difference between them.

\noindent
\textbf{DA Refactoring.} Likewise the former, this analysis also considered the Shapiro-Wilk normality test (\textit{p-value} = 7.69e-06) beforehand. Data normality test indicated no normal distribution, then used the non-parametric Mann-Whitney test. Our results showed a significant statistical difference between the refactoring of DA (\textit{p-value} = 4.764e-05), confirming a significant difference between the refactoring with \texttt{RAIDE} and manual refactoring. Therefore, we refute the null hypothesis (\textit{H\textsubscript{0}2}). The average time was 107.2s (\textit{sd} = 35.86) and 2.55s (\textit{sd} = 1.57s) for manual refactoring and using \texttt{RAIDE}, respectively. Figure \ref{fig:box_comparison} shows that the median line of the box of \texttt{RAIDE} is outside the box of manual refactoring, which confirms a difference between them.

In addition, from a more qualitative standpoint, we also considered the participants' comments on the refactoring with and without \texttt{RAIDE}. P01, P13, and P16 indicated that they would use \texttt{RAIDE} in their projects due to detecting and refactoring the AR and DA test smells. According to P17, \texttt{RAIDE} can increase productivity in the identification and refactoring process, and such resources try to decrease the chance of human error. According to P19, \texttt{RAIDE} is convenient because it visually shows the AR and DA test smells, and their removal is automated, which allows the user to understand the steps taken.\looseness=-1

\begin{tcolorbox}
\textbf{Summary:} The difference between \texttt{RAIDE} and the manual process for refactoring the AR and DA test smells is significant. \texttt{RAIDE} was more than forty-two times faster than the manual process for refactoring the DA test smell. Also, there is a consensus regarding the participants’ opinion about \texttt{RAIDE}. There is an indication that \texttt{RAIDE} makes it easier to refactor the AR and DA test smells.
\end{tcolorbox}

\subsection{Discussion}\label{implicacoes}

After analyzing the gathered data, we could infer that detecting the AR and DA test smells with \texttt{tsDetect} was the most time-consuming task. That result is related to the fragmentation of the \texttt{tsDetect}, which has three intermediary steps until it presents the results. We also highlight the tool's low usability and the laborious process to detect the location of the test smells in the test code. 

The participants' feedback includes essential data we should consider. For example, P01 and P02 reported that although \texttt{tsDetect} treats more test smells than \texttt{RAIDE}, detecting test smells in the former is more cumbersome. They stated that it is necessary to manipulate several files, and the command-line interaction might make it difficult for the identification process. P03 and P06 highlighted that the test smells detection process with \texttt{tsDetect} takes longer than with the \texttt{RAIDE}. P04 claimed that the process performed by \texttt{tsDetect} is counterintuitive, i.e., users would need to repeat the steps more often to learn the process (the order of entry and exit of the files). P06 also mentioned that the results of \texttt{tsDetect} are not very expressive because it only reports the existence or not of test smells, which can lead to misinterpretation, especially when dealing with long test classes. 

Furthermore, the tasks performed with \texttt{RAIDE} took an average of 1.94 minutes, against an average of 14.42 minutes to perform the tasks with \texttt{tsDetect}. In addition, we limited the number of classes and tested the smells of the \textcolor{black}{projects} used to experiment. Therefore, gathered data indicate that manual test code refactoring would take longer in real-world environments, even without \texttt{RAIDE}. Indeed, it is necessary to carry out further studies in this direction to either confirm or refute this statement.

\section{Threats to Validity}\label{section:treats_to_validaty}

\textbf{Internal Validity}. It refers to the effects of the treatments on the variables due to uncontrolled factors in the environment \cite{wohlin2012experimentation}. We used system training to introduce the tools and participants’ tasks to mitigate this threat. We used randomization to assign the order of participants to the tasks to mitigate the learning effect. However, in realizing the assigned tasks, some participants presented difficulties in manually accomplishing the refactoring task (i.e., using the \texttt{tsDetect}). In this case, we guided them to find a solution that may positively influence those participants’ performance using \texttt{tsDetect}.

\textbf{External Validity.} It concerns whether the results can be generalized outside the experimental settings \cite{wohlin2012experimentation}. To mitigate this threat, we counted on experts with different backgrounds. Although there was not a big difference in the number of practitioners and academics, we introduced the detection and refactoring concepts with the tools using a training system. Although the participants had no experience in the context of the experiment, they were skilled enough to perform the tasks.

\textbf{Construct Validity.} It represents the cause and effect concepts to measure in the experiment through dependent and independent variables \cite{wohlin2012experimentation}. In this study, we did not compare \texttt{RAIDE} with a tool that presents the same features. \texttt{tsDetect} tool detects several test smells but does not assist in refactoring them. Conversely, \texttt{RAIDE} detects two test smells and support their refactoring. In the end, we could compare the effects of both manual and automated refactoring. Also, we conducted a pilot study that helped us improve the experiment design and materials.

\textbf{Conclusion Validity.} It refers to the extension of the conclusions about the relationship between the treatments and the outcomes \cite{wohlin2012experimentation}. The main threat to conclusion validity is the small size of the sample. Although we carried out the controlled experiment with 20 participants, we selected most of them with some prior knowledge about test smell.
\section{Related Work} \label{section:related_work}

\citacao{ \cite{greiler2013}} introduced the \texttt{TestHound} tool, which performs a test code static analysis to detect the smells: Dead Field, General Fixture, Lack of Cohesion of Test Method, Obscure In-Line Setup, Test Maverick, and Vague Header Setup. The tool suggests refactoring candidates for removing test smells. The authors conducted a study with users and showed that the tool helps developers understand and adjust the test code. 

\citacao{ \cite{baker2006trex}} proposed \texttt{TRex}, an Eclipse plugin that detects refactoring opportunities to Standardized Tree and Tabular Combined Notation (TTCN-3) test suites. \texttt{TRex} calculates metrics to measure the overall quality of a TTCN-3 test suite and applies a pattern-based analysis to suggest candidate test refactorings. Its accuracy in detecting/refactoring test smells in TTCN-3 test suites, and evaluation with practitioners are unknown.

\citacao{ \cite{Martinez2020}} released \texttt{RTj}, a command-line tool that performs static and dynamic analysis to detect rotten green test smells. \texttt{RTj} also suggests to developers candidate test refactorings. The authors pointed out that \texttt{RTj} detects some false positives due to using conditionals or multiple test contexts. However, there are no details concerning the accuracy of its refactoring capabilities nor experiments validating its usability.

\citacao{ \cite{Lambiase2020}} released \texttt{DARTS} (Detection And Refactoring of Test Smells), a plugin for IntelliJ that utilizes information retrieval to detect three smells (General Fixture, Eager Test, and Lack of Cohesion of Test Methods). They built the tool on top of \texttt{TASTE} \citacao{ \cite{Palomba2018}}, a textual-based detector with an overall f-measure of 67\%, 76\%, and 72\% for the three test smells mentioned above, respectively. The tool also offers refactoring support. As for the refactoring support available in the tool, \texttt{DARTS} exploits the IntelliJ APIs to ensure that the refactored code is compilable and error-free.

Compared with the related tools, \texttt{RAIDE} expands the support to detect and refactor other test smells \cite{santana2020raide}. During plugin development researchers with experience in test smells validated the automated refactorings applied by \texttt{RAIDE}.

\vspace{-3pt}
\section{Conclusion} \label{section:conclusion}

Software test code refactoring is highly dependent on automated support, so it might be cost-effective. Current literature encompasses a few tools supporting automated test code refactoring. However, there is little evidence of automated support to handle test smells refactoring. In prior work, we introduced \texttt{RAIDE}, an Eclipse IDE plugin to automate the test smells detection and refactoring from the JUnit test code. That tool can handle Assertion Roulette and Duplicate Assert test smells in the current version. In this paper, we presented an empirical assessment of the \texttt{RAIDE}. We carried out a controlled experiment with twenty participants to evaluate the tool. The results indicate that \texttt{RAIDE} can detect test smells faster than a comparable state-of-the-art approach. Also, \texttt{RAIDE} was able to refactor a test method in a tiny fraction of time. In particular, the results were very favorable compared to the state-of-the-art approach. Future work directions include extending the \texttt{RAIDE} to consider other test smells and refactoring techniques. Furthermore, there is a need to conduct further empirical studies to validate whether the tool is valuable and effective in real-world practice.

\vspace{-3pt}

\section*{Acknowledgments}

This study was financed in part by the Coordenação de Aperfeiçoamento de Pessoal de Nível Superior - Brasil (CAPES) - Finance Code 001; and FAPESB grants JCB0060/2016 and BOL0188/2020.

\bibliographystyle{sbc}
\bibliography{sbc-template}

\end{document}